\documentclass[preprint]{aastex}

\usepackage{amsmath}
\usepackage{hyperref}
\usepackage{lscape}

\newcommand{\fermi}{\emph{Fermi}}
\newcommand{\DG}{$^{\circ}$}
\newcommand{\mypsr}{PSR B1821$-$24}
\newcommand{\Per}{$P$}
\newcommand{\Pd}{$\dot{P}$}
\newcommand{\Ed}{$\dot{E}$}
\newcommand{\ntoasNRT}{2994}
\newcommand{\mjdminNRT}{1989 October 3 (MJD 47802)}
\newcommand{\mjdmaxNRT}{2012 December 1 (MJD 56262)}
\newcommand{\ntoasWSRT}{81}
\newcommand{\mjdminWSRT}{2004 October 10 (MJD 53288)}
\newcommand{\mjdmaxWSRT}{2012 September 14 (MJD 56184)}
\newcommand{\ntoasJBO}{29}
\newcommand{\mjdminJBO}{2009 August 31 (MJD 55074)}
\newcommand{\mjdmaxJBO}{2012 September 13 (MJD 56183)}
\newcommand{\rmsnonW}{9.2 $\mu$s}
\newcommand{\excursnonW}{17 mP}
\newcommand{\rmsW}{3.1 $\mu$s}
\newcommand{\excursW}{10 mP}
\newcommand{\DMphaseunc}{$\sim$ 1.1 mP}
\newcommand{\Wterms}{eight}

\begin{document}

\title{Broadband Pulsations from PSR B1821$-$24: Implications for Emission Models and the Pulsar Population of M28}

\author{
T.~J.~Johnson\altaffilmark{1,2}, 
L.~Guillemot\altaffilmark{3,4}, 
M.~Kerr\altaffilmark{5,6}, 
I.~Cognard\altaffilmark{7,8}, 
P.~S.~Ray\altaffilmark{9,10}, 
M.~T.~Wolff\altaffilmark{9}, 
S.~B\'{e}gin\altaffilmark{11}, 
G.H.~Janssen\altaffilmark{12}, 
R.~W.~Romani\altaffilmark{5}, 
C.~Venter\altaffilmark{13}, 
J.~E.~Grove\altaffilmark{9}, 
P.~C.~C.~Freire\altaffilmark{3}, 
M.~Wood\altaffilmark{5},
C.~C.~Cheung\altaffilmark{9}, 
J.~M.~Casandjian\altaffilmark{14}, 
I.~H.~Stairs\altaffilmark{15}, 
F.~Camilo\altaffilmark{16,17}, 
C.~M.~Espinoza\altaffilmark{12,18}, 
E.~C.~Ferrara\altaffilmark{19}, 
A.~K.~Harding\altaffilmark{19}, 
S.~Johnston\altaffilmark{20}, 
M.~Kramer\altaffilmark{3,12}, 
A.~G.~Lyne\altaffilmark{12}, 
P.~F.~Michelson\altaffilmark{5}, 
S.~M.~Ransom\altaffilmark{21}, 
R.~Shannon\altaffilmark{20}, 
D.~A.~Smith\altaffilmark{22}, 
B.~W.~Stappers\altaffilmark{12}, 
G.~Theureau\altaffilmark{7}, 
S.~E.~Thorsett\altaffilmark{23}
}
\altaffiltext{1}{National Research Council Research Associate, National Academy of Sciences, Washington, DC 20001, resident at Naval Research Laboratory, Washington, DC 20375, USA}
\altaffiltext{2}{email: tyrel.j.johnson@gmail.com}
\altaffiltext{3}{Max-Planck-Institut f\"ur Radioastronomie, Auf dem H\"ugel 69, 53121 Bonn, Germany}
\altaffiltext{4}{email: guillemo@mpifr-bonn.mpg.de}
\altaffiltext{5}{W. W. Hansen Experimental Physics Laboratory, Kavli Institute for Particle Astrophysics and Cosmology, Department of Physics and SLAC National Accelerator Laboratory, Stanford University, Stanford, CA 94305, USA}
\altaffiltext{6}{email: kerrm@stanford.edu}
\altaffiltext{7}{ Laboratoire de Physique et Chimie de l'Environnement, LPCE UMR 6115 CNRS, F-45071 Orl\'eans Cedex 02, and Station de radioastronomie de Nan\c{c}ay, Observatoire de Paris, CNRS/INSU, F-18330 Nan\c{c}ay, France}
\altaffiltext{8}{email: icognard@cnrs-orleans.fr}
\altaffiltext{9}{Space Science Division, Naval Research Laboratory, Washington, DC 20375-5352, USA}
\altaffiltext{10}{email: Paul.Ray@nrl.navy.mil}
\altaffiltext{11}{D\'{e}partement de physique, g\'{e}nie physique et optique, Universit\'{e} Laval, Qu\'{e}bec, Canada}
\altaffiltext{12}{Jodrell Bank Centre for Astrophysics, School of Physics and Astronomy, The University of Manchester, M13 9PL, UK}
\altaffiltext{13}{Centre for Space Research, North-West University, Potchefstroom Campus, Private Bag X6001, 2520 Potchefstroom, South Africa}
\altaffiltext{14}{Laboratoire AIM, CEA-IRFU/CNRS/Universit\'e Paris Diderot, Service d'Astrophysique, CEA Saclay, 91191 Gif sur Yvette, France}
\altaffiltext{15}{Department of Physics and Astronomy, University of British Columbia, 6224 Agricultural Road, Vancouver, BC V6T 1Z1, Canada}
\altaffiltext{16}{Columbia Astrophysics Laboratory, Columbia University, New York, NY 10027, USA}
\altaffiltext{17}{Arecibo Observatory, Arecibo, Puerto Rico 00612, USA}
\altaffiltext{18}{Departamento de Astronom\'{i}a y Astrof\'{i}sica, Pontificia Universidad Cat\'{o}lica de Chile, Vicu\~{n}a Mackenna 4860, Macul, Santiago, Chile}
\altaffiltext{19}{NASA Goddard Space Flight Center, Greenbelt, MD 20771, USA}
\altaffiltext{20}{CSIRO Astronomy and Space Science, Australia Telescope National Facility, Epping NSW 1710, Australia}
\altaffiltext{21}{National Radio Astronomy Observatory (NRAO), Charlottesville, VA 22903, USA}
\altaffiltext{22}{Centre d'\'Etudes Nucl\'eaires de Bordeaux Gradignan, IN2P3/CNRS, Universit\'e Bordeaux 1, BP120, F-33175 Gradignan Cedex, France}
\altaffiltext{23}{Department of Physics, Willamette University, Salem, OR 97031, USA}


\begin{abstract}
We report a 5.4$\sigma$ detection of pulsed gamma rays from \mypsr\ in the globular cluster M28 using $\sim$44 months of \fermi\ Large Area Telescope (LAT) data that have been reprocessed with improved instrument calibration constants.  We constructed a phase-coherent ephemeris, with post-fit residual RMS of 3 $\mu$s, using radio data spanning $\sim$23.2 years, enabling measurements of the multi-wavelength light curve properties of \mypsr\ at the milliperiod level.  We fold \emph{RXTE} observations of \mypsr\ from 1996 to 2007 and discuss implications on the emission zones.  The gamma-ray light curve consists of two peaks, separated by 0.41$\pm$0.02 in phase, with the first gamma-ray peak lagging the first radio peak by 0.05$\pm$0.02 in phase, consistent with the phase of giant radio pulses.  We observe significant emission in the off-peak interval of \mypsr\ with a best-fit LAT position inconsistent with the core of M28.  We do not detect significant gamma-ray pulsations at the spin or orbital periods from any other known pulsar in M28, and we place limits on the number of energetic pulsars in the cluster. The derived gamma-ray efficiency, $\sim$2\%, is typical of other gamma-ray pulsars with comparable spin-down power, suggesting that the measured spin-down rate ($2.2\times10^{36}$ erg s$^{-1}$) is not appreciably distorted by acceleration in the cluster potential.  This confirms \mypsr\ as the second very energetic millisecond pulsar in a globular cluster and raises the question of whether these represent a separate class of objects that only form in regions of very high stellar density.
\end{abstract}


\section{INTRODUCTION}\label{intro}
Since the launch of the \emph{Fermi Gamma-ray Space Telescope} in 2008, significant high-energy (HE, $\geq$0.1 GeV) pulsations have been detected from more than 40\footnote{See \url{https://confluence.slac.stanford.edu/display/GLAMCOG/Public+List+of+LAT-Detected+Gamma-Ray+Pulsars} for a list of announced gamma-ray pulsars.} millisecond pulsars \citep[MSPs,][mostly in the Galactic field]{2PC} using the Large Area Telescope \citep[LAT, a pair-production telescope sensitive to photons with energies from 20 MeV to $>$300 GeV,][]{LATpaper}, the main instrument aboard \fermi.  
Additionally, HE emission has been detected from the directions of more than a dozen globular clusters \citep{GCpop,Kong10,Tam11}, known or thought to host many MSPs, and the observed spectra of these point sources are consistent with the superposition of emission from several MSPs \citep[predicted by][]{Chen91}.  The one exception is PSR J1823$-$3021A \citep[in the globular cluster NGC 6624,][]{Biggs94} from which significant gamma-ray pulsations have been detected and which accounts for all of the observed HE emission associated with the parent cluster \citep{Freire11}.  To date, all LAT sources associated with globular clusters are consistent with point-like emission, with reported $2\sigma$ upper limits on any extension of $<16$\arcmin\ assuming a two-dimensional Gaussian profile \citep{GCpop}.

MSPs are thought to be old ``recycled'' pulsars that have reached rapid rotation rates via accretion from a companion star \citep[e.g.,][]{Alpar82}.  However, the confirmation of PSR J1823$-$3021A as very energetic MSP suggests an unusual formation scenario, such as the collapse of a white dwarf to a neutron star induced by accretion or a merger with another white dwarf \citep{Ivanova08}, which may be more likely in globular clusters.  As such, it is important to search for and/or confirm similar MSPs in other globular clusters.

Detecting gamma-ray pulsations from more MSPs in globular clusters will help to constrain models for the  broadband emission from the clusters \citep[e.g.,][]{Cheng10,ZBR13,Kopp13}. Constraining the models will generally determine the expected flux level and may be important for extracting the associated particle conversion efficiency from such modeling, once the number of sources is known, thereby constraining the reacceleration particles may undergo within the clusters once they leave the MSP magnetospheres.

\section{PSR B1821$-$24}\label{bio}
Located within the core of the globular cluster M28 (NGC 6626), \mypsr\ is an isolated MSP with a spin period (\Per) of $\sim$3.05 ms and was the first pulsar ever detected in a globular cluster \citep{Lyne87}.  The observed period derivative (\Pd) of $\sim1.62\times10^{-18}$ s s$^{-1}$ \citep{Foster88} leads to an inferred rotational energy-loss rate of $\dot{E} \equiv 4\pi^{2}I\dot{P}/P^{3} \sim2.2\times10^{36}$ erg s$^{-1}$, where $I$ is the moment of inertia of the neutron star and is taken to be $10^{45}$ g cm$^{2}$.  This is the highest \Ed\ of any known rotation-powered MSP, according to version 1.46 of the ATNF Pulsar Database\footnote{\url{http://www.atnf.csiro.au/people/pulsar/psrcat/}} \citep{ATNF}.  While it is possible that the \Pd\ could be artificially enhanced by the gravitational field of the cluster, \citet{Phinney93} showed that the \Pd\ is largely intrinsic. 

\mypsr\ is also the first MSP from which non-thermal pulsed X-ray emission was detected \citep[][using the \emph{Advanced Satellite for Cosmology and Astrophysics}]{Saito97}.  \citet{Rots98} used data from the \emph{Rossi X-ray Timing Explorer} (\emph{RXTE}) and the Green Bank Telescope to determine that the first X-ray and radio peaks were separated by only 0.02 in phase.  Using data from the \emph{Chandra X-ray Observatory}, \citet{Rutledge04} and \citet{Bogdanov11} found that $\sim$15\% of the non-thermal X-ray flux of \mypsr\ was unpulsed.

\mypsr\ was the first MSP ever observed to undergo a glitch \citep{CB04}.  A glitch has also been observed from the mildly-recycled PSR B1913+16 \citep{Weisberg10}.  \citet{RJ01} and \citet{Knight06} reported the detection of giant radio pulses of up to 50 and 91 times the mean pulse intensity, respectively, from \mypsr.  The giant pulses are concentrated in a narrow phase window coincident with the first X-ray peak, similar to what has been observed in the original MSP, PSR B1937+21 \citep{1937GPs}.

Even at a distance ($d$) of $5.1\pm0.5$ kpc \citep[from optical observations of stars in M28,][]{RC91}, the relatively large \Ed\ of \mypsr\ makes it a promising candidate for gamma-ray studies.  
A 4.2$\sigma$ HE pulsed detection was reported by \citet{AGILEpsrs} using data from the \emph{Astro-rivelatore Gamma a Immagini LEggero} (\emph{AGILE}) satellite, but pulsations were only significant in the first 5 days of the observation, the HE pulse profile observed with \emph{AGILE} does not match the LAT profile \citep[see Section \ref{pulses} and][]{wuM28A}, and the observed flux above 100 MeV was greater than the 3$\sigma$ upper limit set using data from the Energetic Gamma-Ray Experiment Telescope \citep{Fierro95}.  The 2FGL catalog \citep{2FGL} associates 2FGL J1824.8$-$2449 with M28 and \citet{GCpop} estimated the number of MSPs in the cluster, based on the HE spectrum, to be 43$^{+24}_{-21}$.  \citet{wuM28A} found a 4.3$\sigma$ pulsed detection using $\sim$42 months of Pass 7 LAT data \citep{P7}, without the updated instrument calibration constants discussed in Section \ref{latdata}, and using the timing solution of \citet{Ray08}, which is not contemporaneous with the LAT data thus leaving open the possibility that the gamma-ray peaks have moved with respect to the radio emission due to timing noise or unmodeled dispersion measure (DM) variations.

\section{OBSERVATIONS AND DATA ANALYSIS}\label{obs}
\mypsr\ is timed under the auspices of the LAT Pulsar Timing Consortium \citep{Smith08}, within which ephemerides are provided from radio observatories around the world for 208 pulsars ranked by $\sqrt{\dot{E}}/d^{2}$.  The timing solution described in Section \ref{radio} will be made available through the \fermi\ Science Support Center\footnote{\url{http://fermi.gsfc.nasa.gov/ssc/data/access/lat/ephems/}}.

\subsection{RADIO TIMING}\label{radio}

The radio timing solution for \mypsr\ has been constructed with the \textsc{Tempo2}\footnote{\url{http://sourceforge.net/projects/tempo2/}} pulsar timing package \citep{T2}, using times of arrival (TOAs) recorded at the Nan\c cay Radio Telescope (NRT) in France, the Westerbork Synthesis Radio Telescope (WSRT) in the Netherlands, and the Lovell Telescope at the Jodrell Bank Observatory in the United Kingdom.

In order to encompass X-ray and \fermi\ LAT observations of \mypsr, we used \ntoasNRT\ TOAs spanning from \mjdminNRT\ to \mjdmaxNRT. Between 1989 October and 2004 November, Nan\c cay pulsar observations were carried out by mixing the signal with a swept frequency local oscillator mimicking the dispersion caused by the interstellar medium, as described in \citet{Cognard1996}; while after late 2004 observations were made using the Berkeley-Orl\'eans-Nan\c cay backend \citep{Nancay}. Although the bulk of radio observations were conducted at 1.4 GHz, the timing data set also included TOAs recorded at different frequencies from 1.6 to 2 GHz, in order to measure and monitor long-term changes in the DM, necessary for comparing profiles at different wavelengths. In addition, a total of \ntoasWSRT\ 1.4 GHz WSRT TOAs recorded with the PuMa and PuMa-II backends \citep{Voute2002,Karuppusamy2008} between \mjdminWSRT\ and \mjdmaxWSRT, as well as \ntoasJBO\ 1.5 GHz Jodrell Bank TOAs \citep{Hobbs04} recorded with the DFB backend between \mjdminJBO\ and \mjdmaxJBO\ were included.

\begin{figure}
\epsscale{0.75}
\plotone{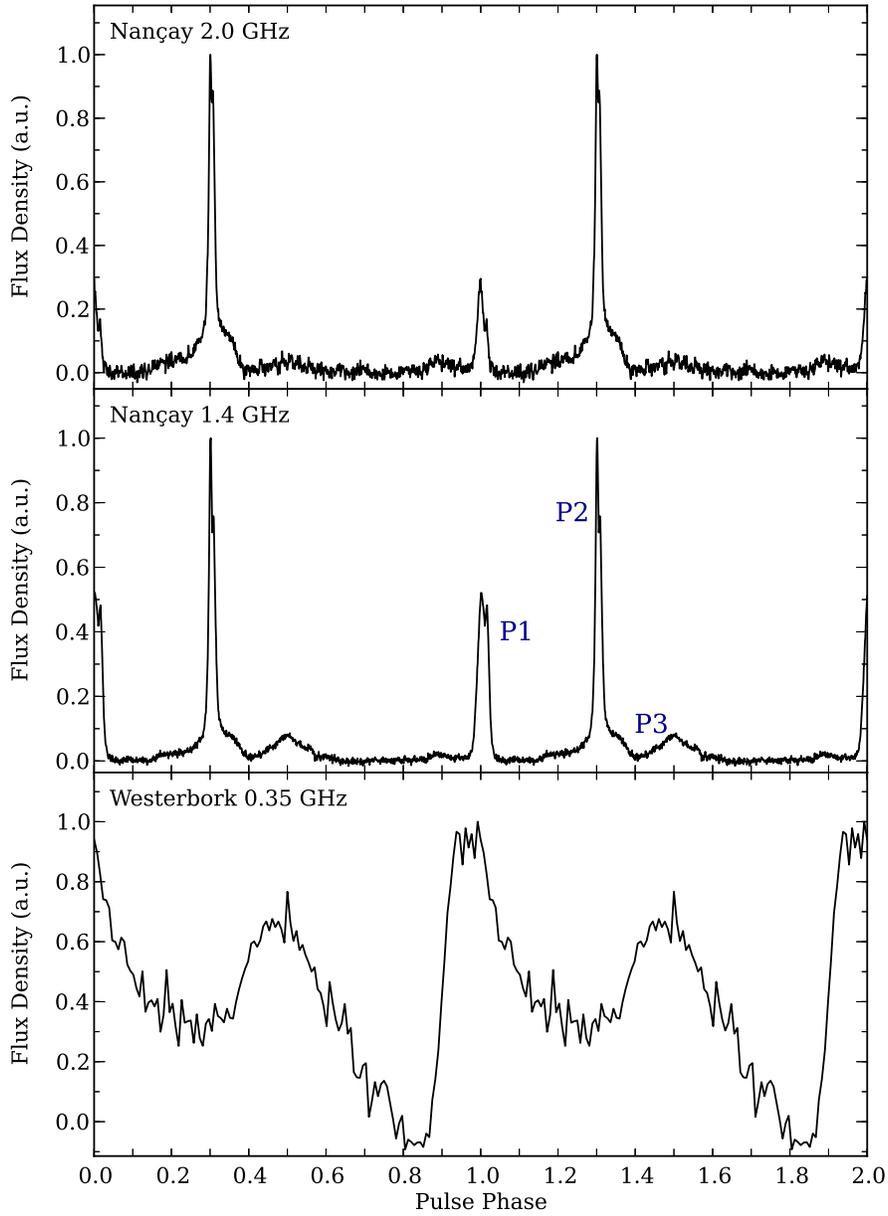}
\caption{Phase-aligned radio light curves of \mypsr, with two pulsar rotations shown for clarity; the peak heights in each panel are normalized independently. From top to bottom: 2 GHz Nan\c{c}ay profile, 1.4 GHz Nan\c{c}ay profile, and 0.35 GHz Westerbork profile.  We denote the second-highest radio peak at 1.4 GHz, near phase 0, as P1; the highest radio peak at 1.4 GHz, near phase 0.3, as P2; and the lowest radio peak at 1.4 GHz, near phase 0.5, as P3.  Both P1 and P3 are also visible at 0.35 and 2 GHz while P2 has no obvious counterpart at 0.35 GHz. \label{fig:radioprofiles}}
\end{figure}

Figure \ref{fig:radioprofiles} shows phase-aligned Nan\c{c}ay radio profiles recorded at 1.4 and 2 GHz, based on $\sim$58.1 hours of observations made between 2008 July 11 and 2011 February 25 and 40.9 hours of observations made between 2004 December 20 and 2008 May 13, respectively; and a 0.35 GHz Westerbork profile obtained by integrating $\sim$8.5 hours of observations conducted between 2013 June 6 and 2013 June 19 with a frequency bandwidth of 0.08 GHz. The relative phase alignment between the 0.35 GHz light curve and the higher frequency radio profiles was estimated by extracting four TOAs from the 0.35 GHz Westerbork data and calculating the average offset between the low frequency and the high frequency Westerbork TOAs with the ephemeris for \mypsr\ obtained from the analysis described below. We estimate that the statistical uncertainty on the relative alignment is on the order of 5 milliperiods (mP). The few 0.35 GHz TOAs were not included in the TOA data set for the timing analysis, having large uncertainties and being affected by strong scattering from the interstellar medium.  For the radio profiles we use the peak naming convention of \citet{BackerSallmen97}, though we shift the first peak to be at phase zero rather than $\sim0.3$.  At 0.35 GHz the P2 is not visible while P1 and P3 appear to broaden and have comparable peak heights.

We first constructed a timing model covering the total TOA data set with good accuracy. At this stage the free parameters were the pulsar position, proper motion, and pulse frequency and first two time derivatives. The published parameters from the glitch in 2001 March \citep{CB04} were included, and refit in the timing model. We then fixed the parameters at the best-fit values and used the Nan\c cay timing data set to determine the DM and its variations. The data set was split into seven intervals spanning two to three years of data, over which TOAs were recorded with a single backend and at multiple frequencies. A DM value was obtained for each of these intervals using \textsc{Tempo2}. A least-squares fit of the seven DM values with a linear function was performed, yielding the values for the DM at MJD 52400 and first time derivative listed in Table \ref{tab:ephem}. The DM and first time derivative were included in the timing model, and frozen at those best-fit values in subsequent analyses. We note that the uncertainty in the DM leads to an uncertainty of \DMphaseunc\ in the conversion of 1.4 GHz TOAs to infinite frequency at the epoch of the ephemeris.

\begin{figure}
\epsscale{0.75}
\plotone{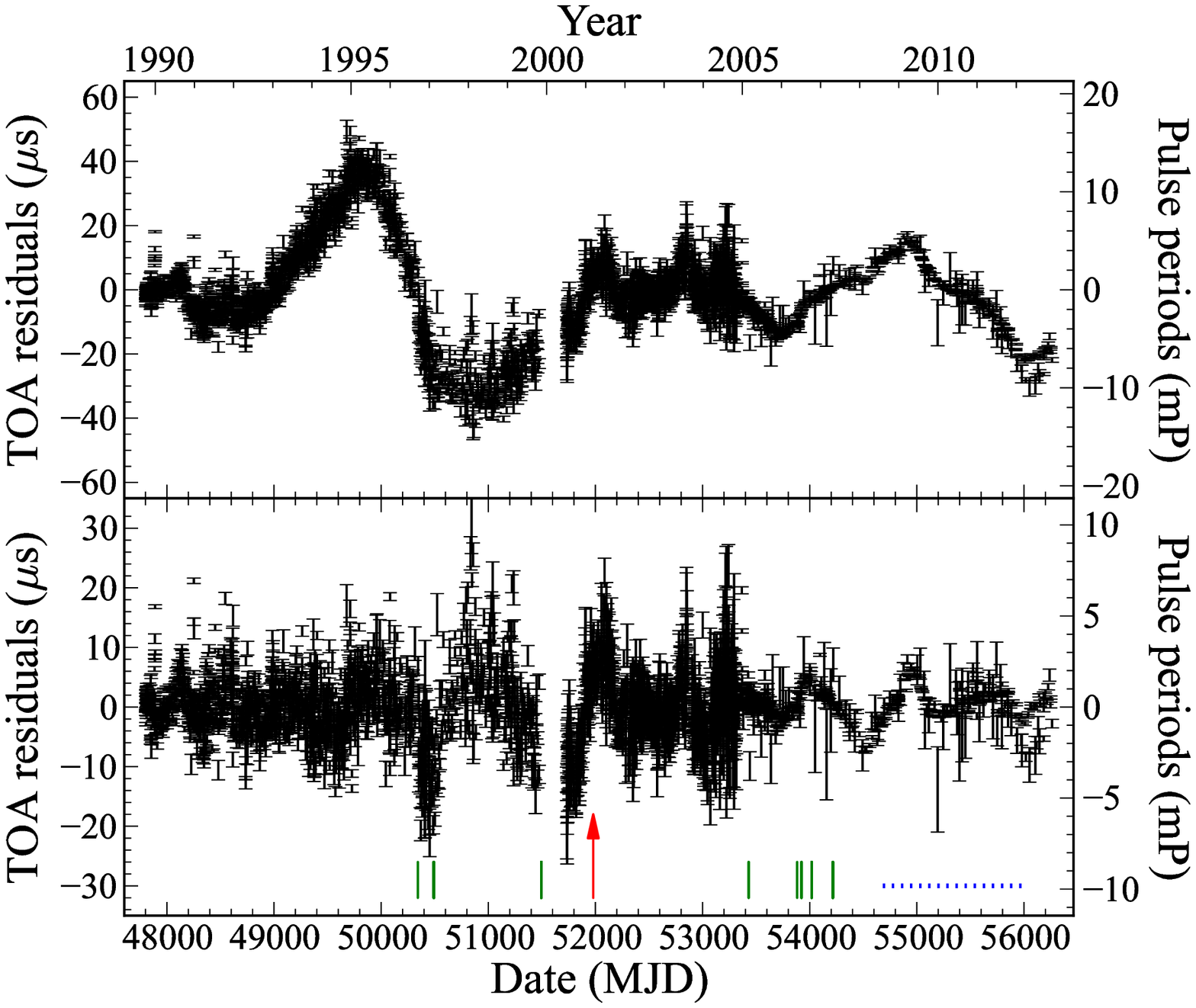}
\caption{Timing residuals as a function of time, for the model given in Table~\ref{tab:ephem} (upper panel), and after whitening of the residuals using \Wterms\ harmonically related sinusoids (lower panel). The arrow (red in the online version) indicates the epoch of the glitch of \mypsr, vertical lines (green in the online version) denote the epochs of the X-ray observations considered in this article, and the dashed horizontal line (blue in the online version) shows the \fermi\ LAT observation interval described in Section \ref{latdata}. \label{fig:residuals}}
\end{figure}

Finally, the timing model was updated by refitting the total TOA data set using the independently-determined DM value and its first time derivative, and leaving other parameters free. The best-fit parameters obtained from this analysis, displayed in Table \ref{tab:ephem}, give an RMS of timing residuals of \rmsnonW, with a maximum excursion of \excursnonW. As can be seen from Figure \ref{fig:residuals}, the TOA residuals exhibit low-frequency structures consistent with rotational irregularities \citep[so-called ``timing noise'', see e.g.,][]{Hobbs04,HLK10}, which we modeled using \Wterms\ harmonically related sinusoids, using the ``FITWAVES'' option of \textsc{Tempo2}, and fixing all other timing parameters. After the whitening procedure, the timing residuals had an RMS of \rmsW, with a maximal excursion of \excursW. The whitened timing residuals are displayed in the lower panel of Figure~\ref{fig:residuals}. The X-ray and gamma-ray timing analyses presented in Sections \ref{xray} and \ref{pulses} were carried out with the whitened timing solution, including the FITWAVES parameters.

The observed \Pd\ of a pulsar can be increased from the true value by contributions from the proper motion \citep{Shklovskii70}.  At a distance of 5.1 kpc and with a total proper motion of 8.5 mas yr$^{-1}$, this effect contributes $\sim2.7\times10^{-21}$ s s$^{-1}$ to the measured \Pd\ of \mypsr, three orders of magnitude less than the value reported in Table \ref{tab:ephem}. Therefore, we do not correct for this effect in the observed and derived parameters of \mypsr.  The latest proper motion measurement for M28 \citep{CD13} agrees well with our values, with a total difference of 21 km s$^{-1}$ at a distance of 5.1 kpc.  This difference is less than the estimated escape velocity of 63.8 km s$^{-1}$ \citep{Gnedin02}, suggesting that \mypsr\ is in fact bound to the cluster.

\begin{deluxetable}{l c}
\tablewidth{0pt}
\tablecaption{Observed and Derived Timing Parameters of \mypsr\ \label{tab:ephem}}
\tablecolumns{2}
\tablehead{\colhead{Parameter} & \colhead{Value}}
\startdata

RA, $\alpha$ (J2000) &  18:24:32.00819(2) \\
Dec., $\delta$ (J2000) &  $-$24:52:10.720(5) \\
Rotational period, \Per\ (s) &  0.00305431496291 \\
First period derivative, \Pd\ ($10^{-18}$ s s$^{-1}$)  &  1.6187747(1) \\
Second period derivative, $\ddot P$ ($10^{-31}$ s s$^{-2}$) &  2.481(5) \\
Proper motion in right ascension, $\mu_\alpha \cos(\delta)$ (mas yr$^{-1}$) & $-$0.25(2) \\
Proper motion in declination, $\mu_\delta$ (mas yr$^{-1}$) &  $-$8.5(4) \\
Epoch of ephemeris (MJD) &  50000 \\
Glitch epoch (MJD) &  51979.5 \\
Glitch frequency step ($10^{-9}$ s$^{-1}$) &  3.0671(7) \\
Glitch frequency derivative step ($10^{-18}$ s$^{-2}$) &  3.156(9) \\
Epoch of dispersion measure determination (MJD) &  52400 \\
Dispersion measure, DM (cm$^{-3}$ pc) &  119.8691(16) \\
Dispersion measure derivative, DM1 (cm$^{-3}$ pc yr$^{-1}$) &  0.0033(2) \\
Span of timing data (MJD) &  47802 --- 56262 \\
Number of TOAs &  3104 \\
RMS of TOA residuals ($\mu$s) &  9.162 \\
Solar system ephemeris model &  DE405 \\
Time system &  TCB \\
\hline
Total proper motion, $\mu_T$ (mas yr$^{-1}$) &  8.5(4) \\
Apparent spin-down luminosity, \Ed\ ($10^{36}$ erg s$^{-1}$) &  2.2 \\
Magnetic field strength at the light cylinder, $B_{\rm LC}$ ($10^5$ G) &  7.2 \\

\enddata
\tablecomments{Measured and derived parameters from the radio observations described in Section \ref{radio}.  Numbers in parentheses are the nominal 1$\sigma$ \textsc{Tempo2} uncertainties in the least-significant digits quoted.}
\end{deluxetable}

\subsection{X-RAY DATA}\label{xray}
The \emph{RXTE} observations we report on here were performed by the Proportional 
Counter Array (PCA, which consists of 5 individual proportional counter units, PCUs) 
from 1996 September 16 (MJD 50342.261) to 2007 April 26 (MJD 54216.252), accumulating a total integration time of $\sim$469 ks. 
These observations employed anywhere from 1 to 5 PCUs in various combinations 
during each observation with data recorded using GoodXenon or GoodXenonwithPropane mode. 
The PCA data were analyzed using the HEASoft version 6.12 data analysis suite.  We employed a variety of bit masks\footnote{\url{http://heasarc.nasa.gov/docs/xte/recipes/cook_book.html}} 
to select events from the PCUs in the 3 to 16 keV range that were on during each individual observation. 
In addition, \citet{Ray08} reported that including events from the first and second anode layer improved the signal-to-noise of the pulsed detection and we followed that prescription here. 
We did not apply a background correction.

The PCA is not an imaging instrument.  Rather, it has a field of view approximately represented by a Gaussian with FWHM of 14\arcmin\ \citep{Jahoda06}.
This means that other X-ray sources known to be in M28 and to have significant flux above 3 keV \citep[e.g.,][]{Becker2003} will contribute to the total count rate in each observation.
Because the contribution from these additional sources will add incoherently to the pulsed signal from \mypsr and we cannot know which events are from \mypsr, we do not attempt to account for these additional X-ray sources in our analysis or to estimate a resulting background level for the pulsed analysis in Section \ref{pulses}. 

The events that satisfy our selection criteria were barycentered with the \emph{faxbary} tool using the DE405 solar system ephemeris and including the \emph{RXTE} fine clock corrections yielding an individual event timing accuracy of $\sim 6\ \mu{\rm s}$ \citep{Rots98,Jahoda06}.
The proper motion of the pulsar was incorporated into the position used to barycenter the data at each epoch.
Pulse phases were calculated utilizing the \texttt{Photon Events} plugin\footnote{Written by Anne Archibald, \url{http://www.physics.mcgill.ca/~aarchiba/photons\_plug.html}} 
for \textsc{Tempo2} and the radio ephemeris described in Section \ref{radio}.

\subsection{LAT DATA: P7REP}\label{latdata}
Pass 7 LAT data have been reprocessed\footnote{For more information about the updated calibrations and P7REP data see \citet{P7REP} and \url{http://fermi.gsfc.nasa.gov/ssc/data/analysis/documentation/Pass7REP_usage.html }.} using updated calibration constants for the detector subsystems, most importantly for the calorimeter (CAL) to more accurately describe the position-dependent response of each scintillator crystal and the slight decrease in scintillation light yield with time ($\sim$1\% per year) from radiation exposure on orbit.

This reprocessing affected the LAT data (P7REP, hereafter) in several ways.  First, the point-spread function (PSF) is significantly improved above a few GeV, with a reduction in the 68\% containment radius of 30\% (40\%) for events converting in the front (back) of the tracker \citep{P7REP}.  At these energies, the improved calibration constants result in more accurately calculated centroids of energy deposition in the CAL to constrain the incident event direction.  Second, the significance of detection and precision of measured photon flux is increased slightly for most sources -- more strongly for sources with hard spectra than for those, like pulsars, with cutoffs at a few GeV.  Third, spectral features such as cutoff energies are shifted upward slightly in energy ($\sim$few \%) by the change in energy scale.

We selected events from the P7REP data corresponding to the \texttt{SOURCE} class recorded between 2008 August 4 and 2012 March 31; with reconstructed directions within 11$\fdg$5 of the pulsar radio position, allowing us to construct a $16^{\circ}\times16^{\circ}$ square region with no blank corners for a binned likelihood analysis (see Section \ref{spec}); energies from 0.1 to 100 GeV, the lower limit is that recommended for analysis of P7REP data and the upper limit adequately covers the range of known pulsar cutoff energies; and zenith angles $\leq$100\DG, to reduce contamination of gamma-rays from the limb of the Earth.  Good time intervals were then selected corresponding to when the instrument was in nominal science operations mode, the rocking angle of the spacecraft did not exceed 52\DG, the limb of the Earth did not infringe upon the region of interest, and the data were flagged as good.  All LAT analyses were performed using the \fermi\ Science Tools v9r27p1.

The recommended instrument response functions (IRFs, which include the PSF, effective area, and energy dispersion) for analyzing P7REP data are \texttt{P7REP\_V15}.  These IRFs are derived from detailed simulations of the instrument \citep{P7} with some modifications based on on-orbit performance checks, which are detailed below.

The accuracy with which incoming event directions are reconstructed is dependent on the energy ($E$), interaction point within the instrument, and angle with respect to the boresight\footnote{For more details see \url{http://www.slac.stanford.edu/exp/glast/groups/canda/lat\_Performance.htm}\\and \citet{P7}.} ($\theta$).  For a SOURCE class event converting in the front of the instrument, the energy-dependent 68\% confidence-level containment radius, averaged over the acceptance, can be approximated as $\Theta_{68}(E)= \sqrt{(0\fdg66(E/1\ \rm{GeV})^{-0.76})^{2}+(0\fdg08)^{2}}$.

Although the reprocessing significantly improved the PSF at high energies, the angular distribution of gamma rays around point sources used for in-flight calibration of the PSF above 3 GeV was still found to be slightly broader in the P7REP data than predicted by the Monte Carlo (MC) PSF.  The on-orbit PSF for the \texttt{P7REP\_V15} IRFs was derived by rescaling the MC PSF to match the angular distribution of gamma rays around the Vela pulsar below 10 GeV and a sample of bright, high-latitude blazars above 10 GeV.  This correction to the MC PSF model rescales the size of the PSF as a function of energy while preserving the dependence on $\theta$; formerly, for the \texttt{P7\_V6} IRFs, recommended for analyzing the original Pass 7 data, the $\theta$ dependence was not preserved in making this correction \citep{LATPSF}.

There is a known discrepancy between the fluxes arising from analyses using only events that convert in the front or the back of the tracker subsystem \citep[see Figure 47 and Section 5.6 of][]{P7}.  This discrepancy occurs mainly at energies below 300 MeV with differences of $\lesssim10$\%.  The \texttt{P7REP\_V15} effective area tables include an empirical correction for this that does not modify the overall effective area inferred from MC studies.  The total effective area for a near on-axis, 1 GeV, SOURCE class gamma ray is $\sim$7000 cm$^{2}$.

\section{GAMMA-RAY RESULTS}\label{gamma}

\subsection{SPECTRAL AND SPATIAL ANALYSIS}\label{spec}
A binned maximum likelihood analysis was performed on a $16^{\circ}\times16^{\circ}$ region centered on the pulsar position using the \texttt{P7REP\_SOURCE\_V15} IRFs.  All sources from a three-year source list, produced following the same procedure used for the 2FGL catalog, using the original Pass 7 data, and \texttt{P7SOURCE\_V6} IRFs, within 15\DG\ of \mypsr\ were included in the model of the region and all spectral parameters of sources within 8\DG\ (23 point sources and 2 extended sources) were left free.  The Galactic diffuse emission was modeled using the \textit{gll\_iem\_v05.fit} model while the isotropic diffuse emission and residual instrument background were jointly modeled using the \textit{iso\_source\_v05.txt} template\footnote{The P7REP data, \texttt{P7V15 IRFs}, and diffuse models will be available for download at \url{http://fermi.gsfc.nasa.gov/ssc/data/}.}.  These diffuse models were produced specifically for the P7REP data using a refined approach in which residuals in the LAT data were used to fit components of the diffuse emission not derived from observations at other wavelengths \citep[see][]{JeanICRC}.

We modeled the spectrum of \mypsr\ as both a simple power law (Eq.~\ref{pl}) and an exponentially-cutoff power law (Eq.~\ref{ecpl}).
\begin{equation}\label{pl}
\frac{dN}{dE}\ =\ N_{0}\ \Big(\frac{E}{1\ \rm{GeV}}\Big)^{-\Gamma}
\end{equation}
\begin{equation}\label{ecpl}
\frac{dN}{dE}\ =\ N_{0}\ \Big(\frac{E}{1\ \rm{GeV}}\Big)^{-\Gamma}\ \exp\Big\lbrace-\frac{E}{E_{\rm C}}\Big\rbrace.
\end{equation}
\noindent{}Using the likelihood ratio test, a simple power-law shape is ruled out, in favor of an exponentially-cutoff power law, with a confidence level of 5.6$\sigma$.  We detect a point source at the position of \mypsr\ with a likelihood test statistic \citep[TS,][]{2FGL} of 438.  The best-fit spectrum has $E_{\rm C} = 6.1\pm2.1$ GeV, $\Gamma = 2.2\pm0.1$, and gives integral photon and energy fluxes (from 0.1 to 100 GeV) of $F=(7.2\pm0.9)\times10^{-8}$ cm$^{-2}$ s$^{-1}$ and $G=(3.8\pm0.3)\times10^{-11}$ erg cm$^{-2}$ s$^{-1}$, respectively, all uncertainties being statistical.  \mypsr\ is a relatively faint source for the LAT and statistical uncertainties in these measurements dominate the systematic uncertainties; thus, we do not attempt to estimate systematic uncertainties on the best-fit parameters.

While we kept this point source fixed to the radio timing position of \mypsr\ in the spectral analysis, the best-fit LAT position (using the \fermi\ Science Tool \texttt{gtfindsrc}) is right ascension (J2000) = 18:24:43.2 , declination (J2000) = $-$24:51:36.0, with a 95\% confidence-level error radius $r_{95}$ = 4\arcmin 12\arcsec\, which is 2\arcmin 24\arcsec\ from the timing position.

The 2FGL catalog and \citet{wuM28A} have both reported flux values for point sources associated with \mypsr\ using the original Pass 7 data and the \texttt{P7SOURCE\_V6} IRFs in the 1 to 100 GeV and 0.2 to 300 GeV energy ranges, respectively.  Integrating our phase-averaged results over the same energy ranges yields higher values than reported by those authors, by on the order of 20\%.  These differences are larger than expected from switching to P7REP data alone.  We note that the disagreement with the 2FGL flux is at the 2$\sigma$ level and is likely just statistical fluctuation, while the disagreement with \citet{wuM28A} is $<1\sigma$.

We repeated the analysis described in \citet{wuM28A} using similar time, energy, and angular selections and the original Pass 7 data; with the same 2FGL point sources free and fixed in our model of the region; and with the same diffuse components.  However, we found values more consistent with results from our analysis, described previously.  Additionally, our re-analysis only found a TS of 248 for a point source at the position of \mypsr, much less than the value of 825 reported by \citet{wuM28A}.  We note that \citet{2FGL} reported a significance of $\sim11\sigma$ for 2FGL J1824.8$-$2449, using two years of data, which corresponds to a TS of $\sim$144.  Extrapolating to $\sim$42 months we expect a TS of $\sim$200, for a non-variable source, which agrees with our re-analysis when accounting for differences in event selection.  While the differences in $\Gamma$ and $E_{\rm C}$ may be related to the choice of minimum energy and differences in the diffuse model, the disagreement between the TS values is not understood.

Using the initial phase-averaged results, we were able to detect significant pulsations ($>5\sigma$, see Section \ref{pulses}) from \mypsr, the gamma-ray light curve is characterized by two peaks at phases of $\sim$0.0 and $\sim$0.5, similar to the results of \citet{wuM28A}.   However, there was a clear offset above the estimated background level observed in the gamma-ray light curve.  While it is possible that \mypsr\ has a near 100\% duty cycle \citep[as seems to be the case for PSR J1836+5925,][]{LATJ1836} we performed an analysis of the off-peak phase interval (defined to be $\phi\in(0.24,0.34)\cup(0.58,0.82)$) to study the emission in more detail.

We first attempted to ascertain if this emission could be attributed to any of the other known pulsars in M28\footnote{\url{http://www.naic.edu/~pfreire/GCpsr.html}} \citep[11 MSPs and one young, non-recycled pulsar,][and B\'{e}gin et al.~in preparation]{Bogdanov11}.  At a distance of 5.1 kpc, it is possible that the combined emission from these and any other unknown pulsars, less that of \mypsr, may account for the observed off-peak emission.  We obtained timing solutions for PSRs J1824$-$2452B-L (detailed in B\'{e}gin et al.~in preparation) and searched for a periodic signal from each pulsar, at the spin and orbital periods, in the LAT data using event weights \citep[a probability for each event to have originated from the source of interest based on the spectral and spatial model of the region,][]{Kerr11} calculated from the initial phase-averaged analysis.  We used both the full data set and the off-peak interval but found no signal with more than $2\sigma$ significance.

Using the off-peak interval, the best-fit LAT position for this emission is right ascension (J2000) = 18:25:02.4, declination (J2000) = $-$24:43:48.0, with $r_{95}$ = 6\arcmin.  This position is 11\arcmin 24\arcsec\ from the core of M28, nearly twice $r_{95}$.  All of the other known pulsars in M28 are within $\lesssim18$\arcsec\ of \mypsr\ except for J1824$-$2452F which is 2\arcmin\ 45.6\arcsec\ away but still inconsistent with the off-peak emission ($\sim1.5\ r_{95}$ away).  Our model of the region includes only one other point source within $1\fdg5$ of the timing position of \mypsr.  This source has an integral flux, from 0.1 to 100 GeV, of $\sim0.9\times10^{-8}$ cm$^{-2}$ s$^{-1}$ and a photon index of $\sim2$.  There is one additional source within 3\DG\ of \mypsr\ with an integral flux, from 0.1 to 100 GeV, of $\sim4.5\times10^{-8}$ cm$^{-2}$ s$^{-1}$ and a photon index of $\sim2.5$.  All other sources are $>3\fdg5$ from \mypsr.  Therefore, the localization of the off-peak emission should not be strongly affected by known nearby sources.

\begin{figure}
\mbox{\includegraphics[width=0.5\textwidth]{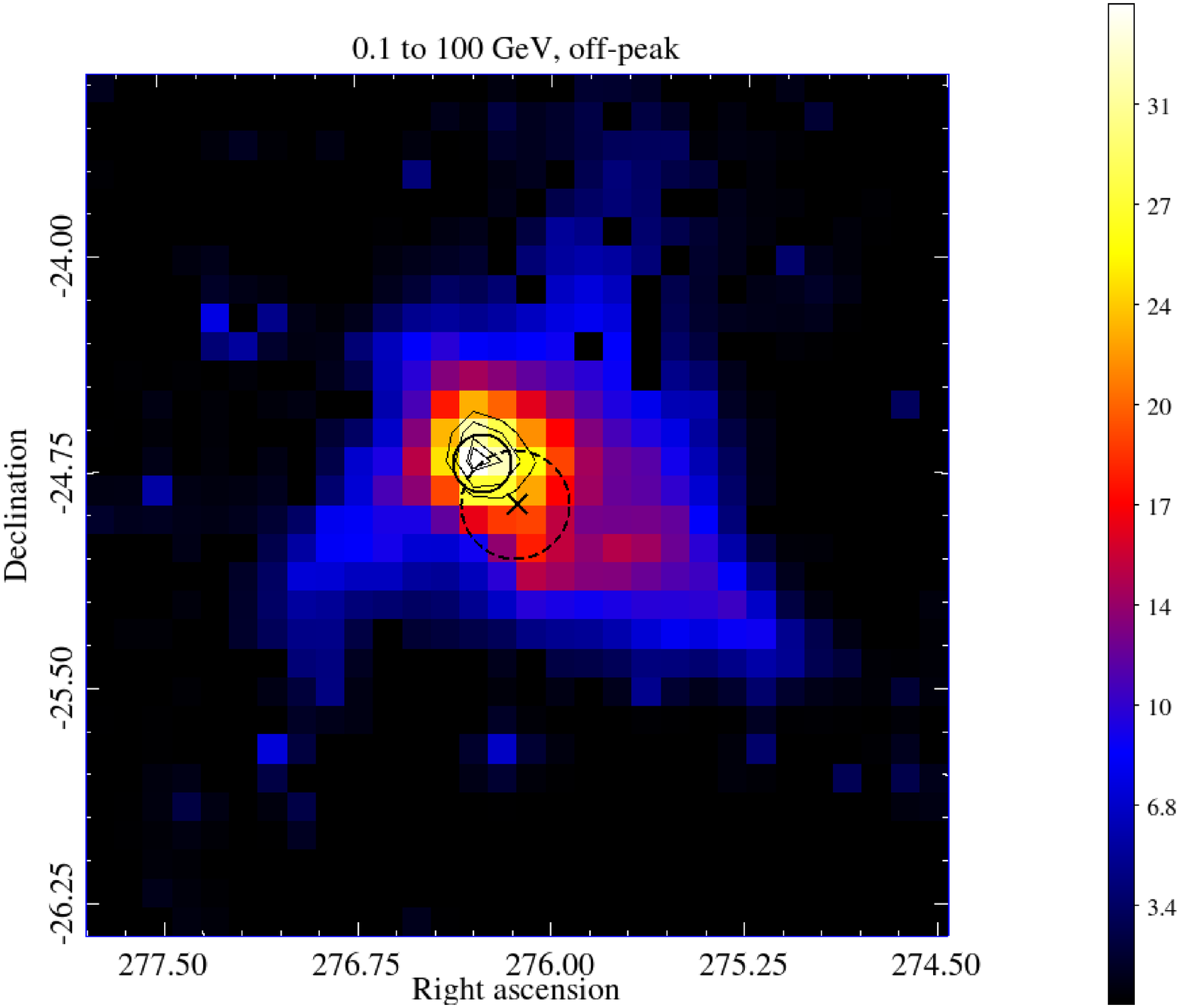}}
\mbox{\includegraphics[width=0.5\textwidth]{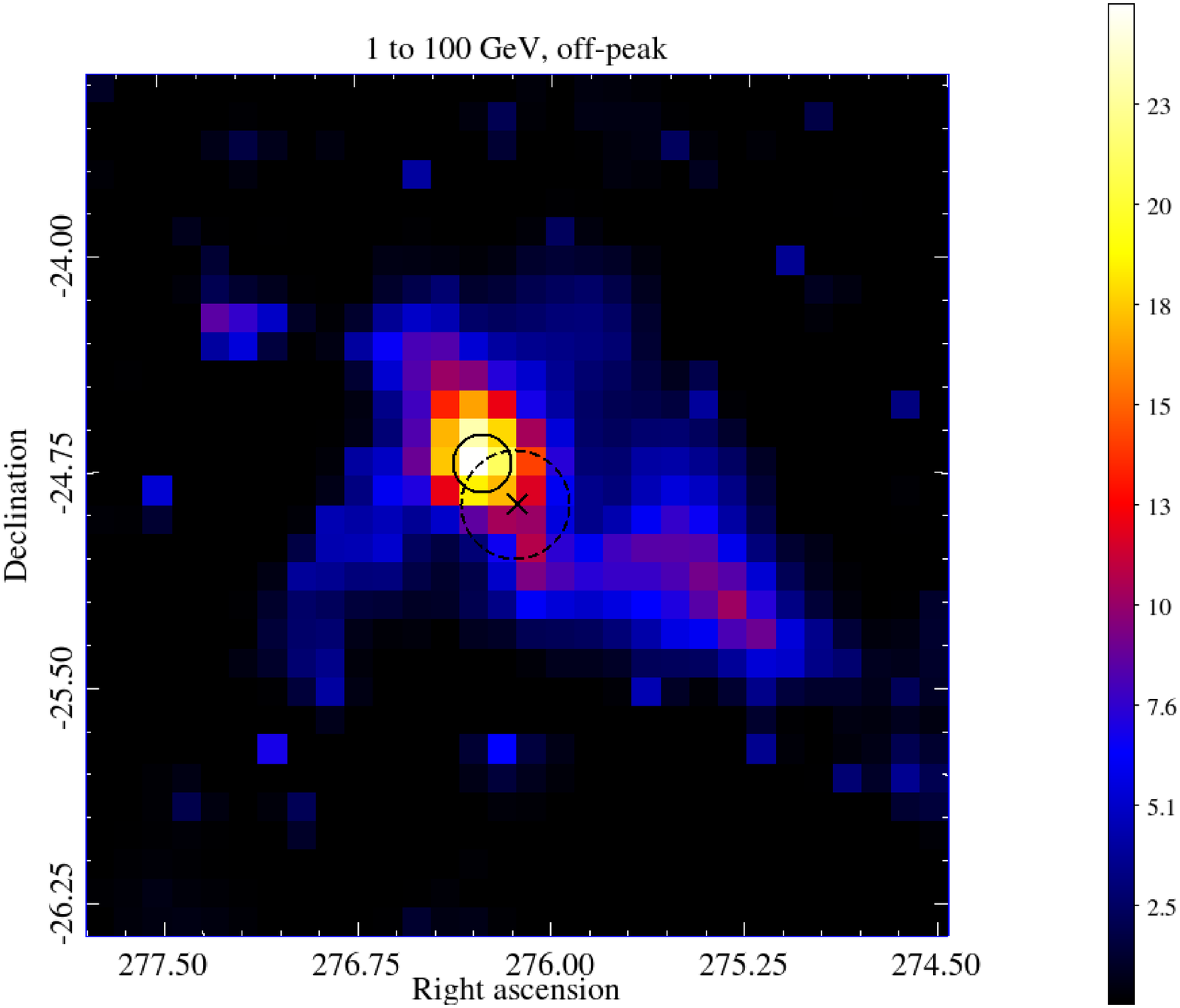}}
\mbox{\includegraphics[width=0.5\textwidth]{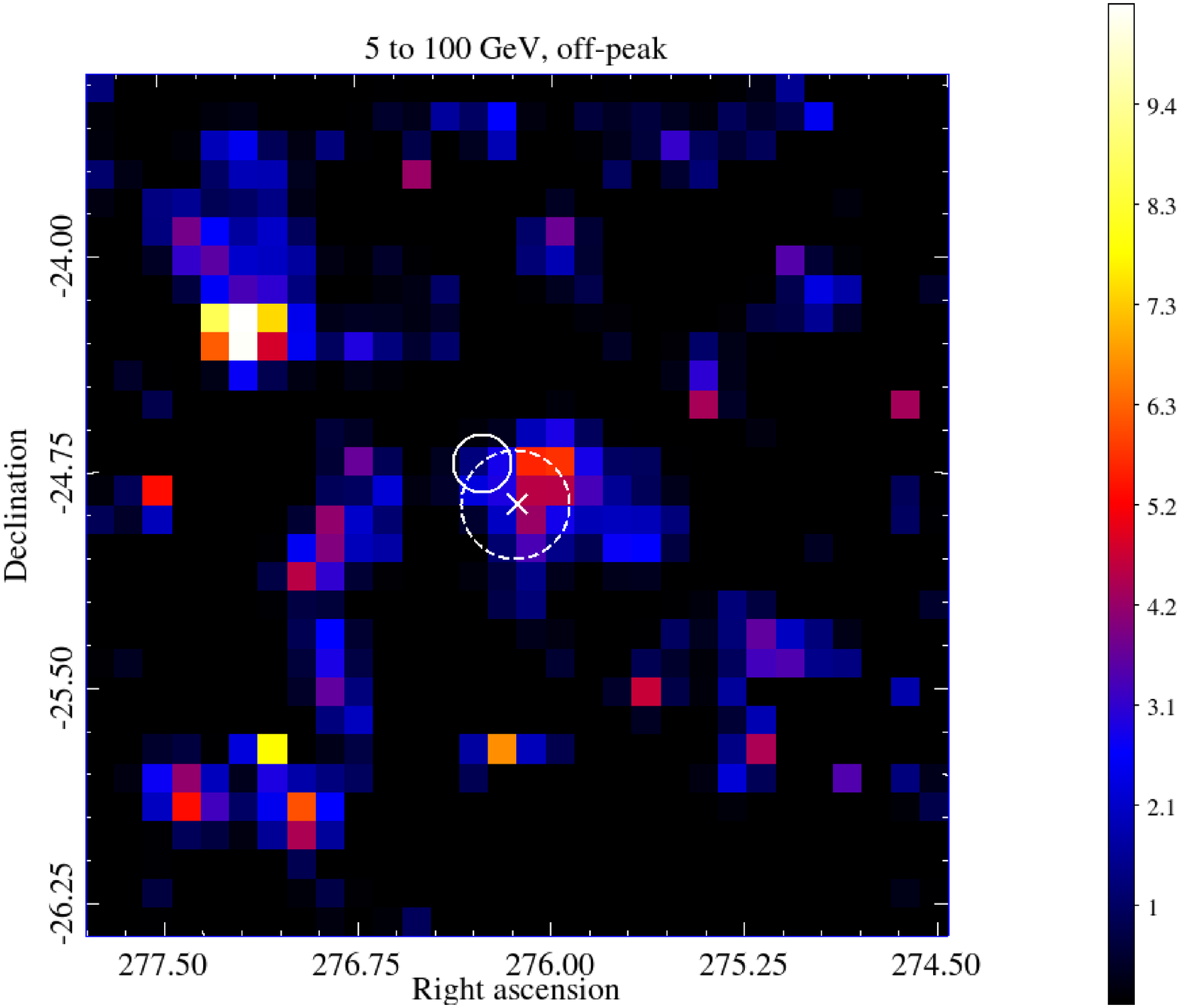}}
\mbox{\includegraphics[width=0.5\textwidth]{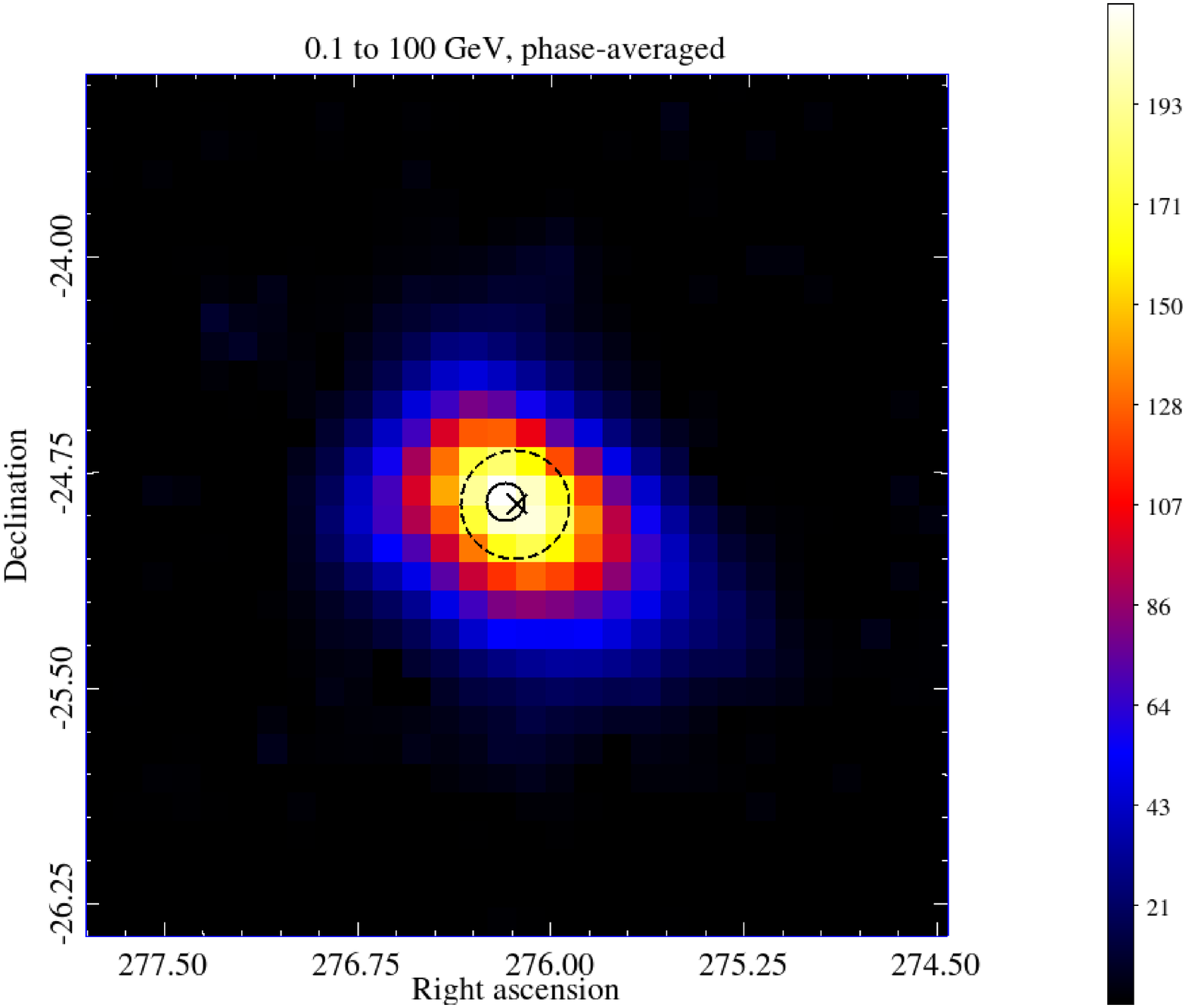}}
\caption{TS maps in the off-peak interval of \mypsr\ using events from 0.1 to 100 GeV (\emph{top left}), 1 to 100 GeV (\emph{top right}), and 5 to 100 GeV (\emph{bottom left}) and the entire phase interval from 0.1 to 100 GeV (\emph{bottom right}).  The TS maps cover $3^{\circ}\times3^{\circ}$, centered on the timing position of \mypsr, and have pixels $0\fdg1$ on a side.  The timing position of \mypsr\ is indicated by the X-point, the best-fit LAT position and positional uncertainty from the corresponding phase interval by the solid circle, and the tidal radius of M28 by the dashed circle. The thin black contours in the top left panel represent the 50, 68, 95, and 98\% confidence levels.\label{fig:offTSmaps}}
\end{figure}

To verify the \texttt{gtfindsrc} position, we built TS maps in the off-peak interval with different minimum energies and a $3^{\circ}\times3^{\circ}$ region centered on the pulsar (using the \fermi\ Science Tool \texttt{gttsmap} in binned mode, see Figure \ref{fig:offTSmaps}).  These maps are constructed by calculating the TS value of a hypothetical point source, with a power-law spectral model, at a grid of positions (constructed by dividing the region into pixels $0\fdg1$ on a side).  While there may be some residual emission associated with M28, the peaks of the TS maps agree well with the best-fit position, except for the TS map above 5 GeV for which we find no significant TS at any position.  The $\Delta$TS contours of the 0.1 to 100 GeV TS map agree well with the off-peak $r_{95}$ from \texttt{gtfindsrc}.  Spectral analysis of the off-peak emission shows no evidence for a cutoff in the spectrum; a power-law fit yields $\Gamma = 2.5\pm0.1$ with $F=(6.7\pm1.1)\times10^{-8}$ cm$^{-2}$ s$^{-1}$ and $G=(3.0\pm0.2)\times10^{-11}$ erg cm$^{-2}$ s$^{-1}$, where the flux values have been rescaled to the full phase interval.

Within the LAT $95 \%$ confidence-level error circle of the off-peak emission, we found no cataloged NVSS \citep{Condon98} radio or RASS \citep{Voges00} X-ray sources down to the typical flux limits of $\sim2.5$ mJy (1.4 GHz) and $\sim3 \times 10^{-13}$ erg cm$^{-2}$ s$^{-1}$ ($0.1-2.4$ keV) of the respective surveys. The lack of a bright radio / X-ray source, combined with the steep LAT gamma-ray spectrum makes a background blazar counterpart unlikely \citep[see][]{BlazarSEDs}.  The Sun does pass close to M28 and is a significant and persistent source of HE gamma rays \citep{SunPaper}; however, the off-peak emission is at an ecliptic latitude of approximately $-1\fdg4$, which is sufficiently offset from the ecliptic plane to rule out an association with the Sun.  As can be seen in Figure \ref{fig:offTSmaps}, the error circle is still consistent with the tidal radius of M28 \citep[11.27\arcmin,][]{Trager95,Chun12} so we cannot completely rule out an association with the cluster, but the interpretation of this emission as the combination of unresolved pulsars is uncertain unless there is a systematic shift in the best-fit localization.  PSR J1824$-$2452F is several core radii away from the center of M28, providing some evidence for the possibility of pulsar ejection from the center of the globular cluster.  Therefore, it is possible that the off-peak emission is an energetic pulsar that has been ejected from M28.  However, the lack of spectral curvature in the off-peak emission (cutoff only preferred at the 1.5$\sigma$ level) might argue against such an interpretation.

Under the hypothesis that the off-peak emission described above is not associated with M28, we performed a spectral analysis in the off-peak interval with a source at the position found previously (not consistent with the cluster) and with a source at the position of M28.  The M28 source is found with a TS of 0.05, which is not significant.  Therefore, we calculated 95\% confidence-level upper limits on the integral photon and energy fluxes from the direction of M28 in the off-peak interval of $F\leq6.3\times10^{-9}$ cm$^{-2}$ s$^{-1}$ and $G\leq7.0\times10^{-12}$ erg cm$^{-2}$ s$^{-1}$, assuming a power-law spectral model with $\Gamma=2$.  We find no evidence for significant flux variability in the off-peak emission but do note a possible slow rise in the flux on 6-month to 1-year timescales.

We repeated the phase-averaged analysis with the off-peak source included in the model, at the best-fit position and with all spectral parameters fixed.  We find a point source at the position of \mypsr\ with TS = 76.  A simple power-law model is rejected in favor of an exponentially-cutoff power-law model at the 3.9$\sigma$ level.  The best-fit spectrum yields $E_{\rm C} = 3.3\pm1.5$ GeV, $\Gamma = 1.6\pm0.3$, and integral fluxes of $F=(1.5\pm0.6)\times10^{-8}$ cm$^{-2}$ s$^{-1}$ and $G=(1.3\pm0.2)\times10^{-11}$ erg cm$^{-2}$ s$^{-1}$.  Given the disagreement between the location of the off-peak emission and the timing position of \mypsr, we consider these values, rather than those from the initial phase-averaged analysis, to best represent the spectrum of the pulsar.

\begin{figure}
\epsscale{0.75}
\plotone{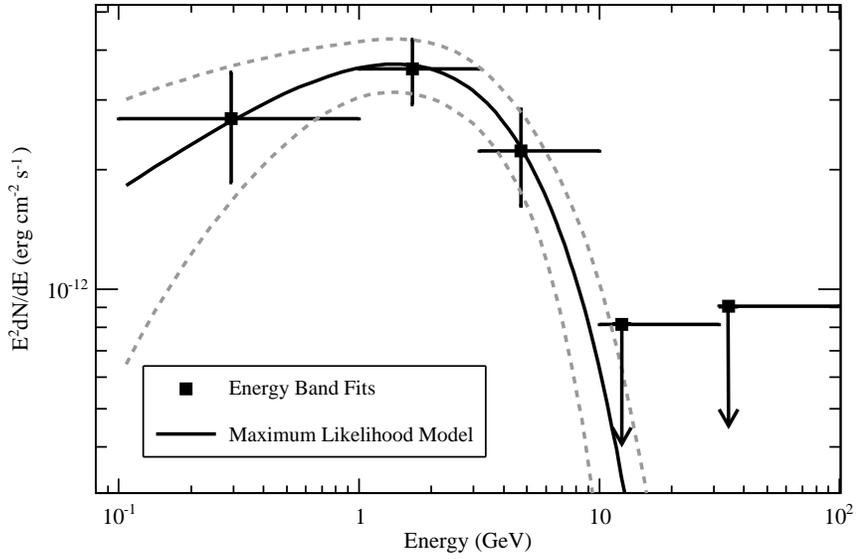}
\caption{Phase-averaged gamma-ray spectrum of \mypsr\ with the off-peak source included in the model.  The black line shows the best-fit model from the likelihood fit over the full energy range; dashed lines show the 1$\sigma$ confidence region.  The pulsar was assumed to have a power-law spectrum in each energy band and required to be found with a TS of at least 9 or else a 95\% confidence-level upper limit was calculated.\label{fig:psrSpec}}
\end{figure}

The gamma-ray spectrum of \mypsr\ is shown in Figure \ref{fig:psrSpec}.  The flux points are derived from fits to the indicated energy bands in which the spectrum of \mypsr\ was modeled as a power law with $\Gamma$ fixed to 2.  The center of each bin is the weighted average energy using the spectral shape of the full energy range fit as the weights.  This leads to the center energies moving closer to the low side of each bin with increasing energy since the pulsar is modeled with a cutoff in the full energy range fit.  We required the source to be detected with a TS of at least 9 ($\sim3\sigma$ for 1 degree of freedom) or else a 95\% confidence-level upper limit on the flux was calculated.

\clearpage

\subsection{PULSATIONS}\label{pulses}
\begin{figure}
\epsscale{0.8}
\plotone{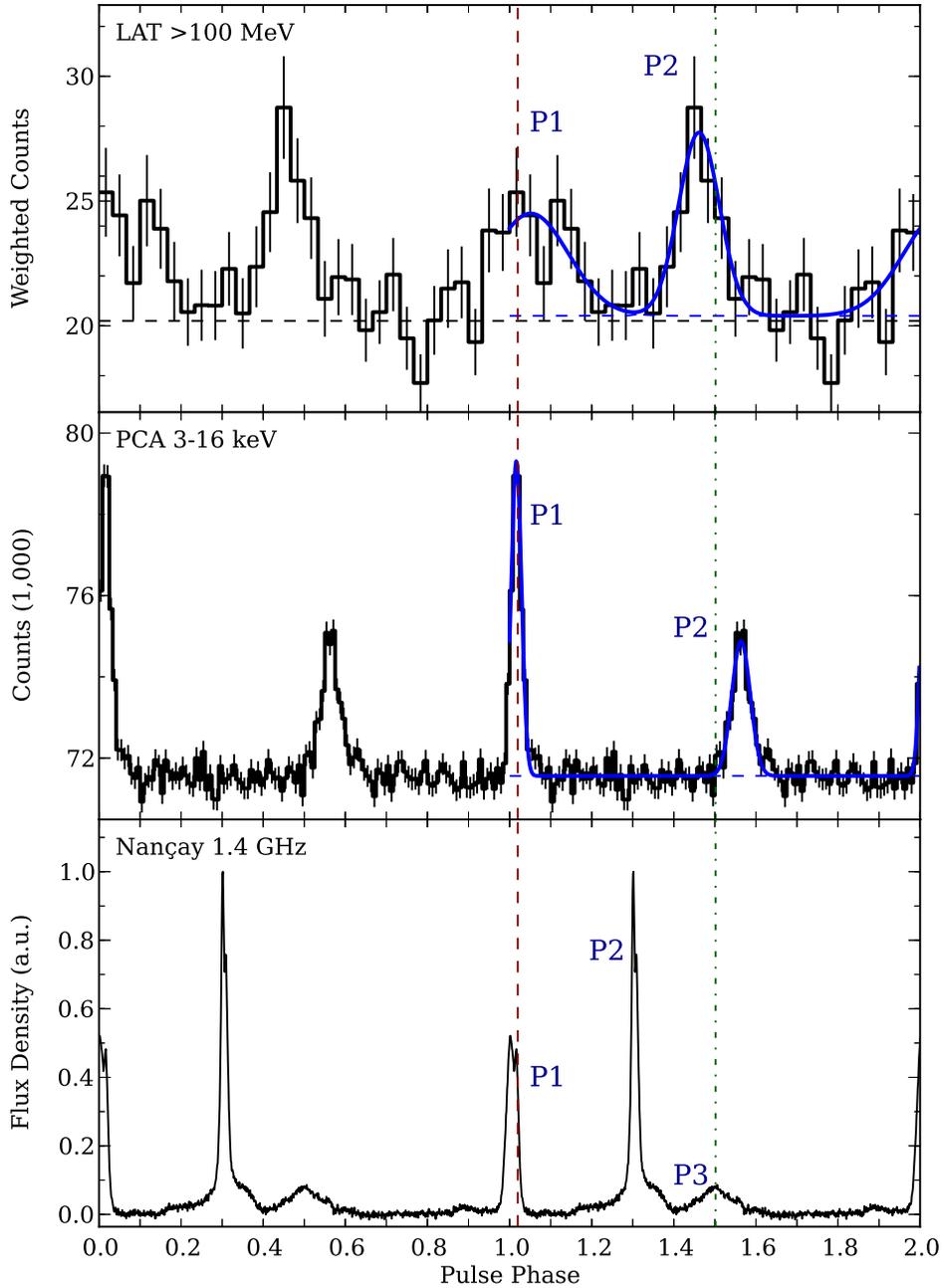}
\caption{Folded light curves of \mypsr, from top to bottom: $\geq$100 MeV, 3 to 16 keV, and 1.4 GHz.  The light curves are shown over two rotations for clarity, the solid (blue in the online version) lines over the second rotation in the top two panels are the best-fit light curve shapes.
The dashed (red in the online version) vertical line indicates the approximate phase from which giant pulses have been observed.  The dot-dash (green in the online version) vertical line indicates the center of P3 in the radio profiles.\label{fig:lc}}
\end{figure}

We selected events with reconstructed directions within 2\DG\ of \mypsr\ and used our best-fit, phase-averaged spectral model, with the off-peak source included in the model, to calculate a probability for each event to be associated with \mypsr.  Events triggering the LAT are time stamped using an on-board GPS receiver that is accurate to within $<$1 $\mu$s relative to UTC \citep{OnOrb}.  We then folded the events at the radio period using the \texttt{fermi} \textsc{Tempo2} plugin \citep{Ray11} and calculated the spectrally-weighted H-test significance \citep{Kerr11}, resulting in a 5.4$\sigma$ pulsed detection.  The light curves of \mypsr\ at different wavelengths are shown in Figure \ref{fig:lc}.  The uncertainties for each bin of the gamma-ray light curve and the background level are calculated as described in \citet{Guillemot12}.  This confirms the periodic signal candidate reported by \citet{wuM28A} and firmly establishes \mypsr\ as a gamma-ray pulsar.

We used photon-weighted maximum likelihood \citep{2PC} to fit parametric functions (light curves) to the LAT and \emph{RXTE} data.  The gamma-ray light curves were fit using an unbinned analysis.  The X-ray event phases were binned into 1000 bins, yielding time resolution comparable to that of the radio ephemeris.  For a set of event phases and weights ($\phi_i$ and $w_i$) this likelihood is given by $\log\mathcal{L} = \sum_{i=0}^{N_{\gamma}} \log \left((1-w_i) + w_i\,f(\phi_i,\psi)\right)$, where $f(\phi,\psi)$ is the assumed functional form with parameters $\psi$.  We fit each peak of the gamma-ray and X-ray data with a symmetric Gaussian shape, because asymmetric peaks were not significantly preferred by the likelihood, and report the best-fit values in Table \ref{lcfits}.  We considered Lorentzian shapes for each peak but found comparable likelihood values and, thus, report only results of the Gaussian fits.  The weights for gamma-ray events are from the phase-averaged spectral fit, while we set $w_i=1$ for \emph{RXTE} data.  For the X-ray and gamma-ray light curves we identify peaks 1 and 2 in the order they appear in phase (as labeled in Figure \ref{fig:lc}).  Using these fits, the first gamma-ray peak spans the phase range $\phi\in[0.0,0.23]\cup[0.87,1.0)$ and the second spans the phase range $\phi\in[0.36,0.56]$, where the quoted ranges correspond to the peak positions plus and minus twice the best-fit widths.

\citet{RJ01} and \citet{Knight06} reported that the first X-ray peak was consistent with the phase at which giant pulses were observed in the radio ($\sim0.02$ in phase after the first radio peak).  While the phases of the first X-ray and gamma-ray peaks are not consistent with 0.02 within uncertainties, we note that 0.02 is only an estimate and thus confirm that the first X-ray peak and now the first gamma-ray peak are consistent with the phase of giant pulses.  \citet{Knight06} also observed a single giant pulse occurring 0.55 in phase after the bulk of the giant pulses, which they contend represents a second population of giant pulses from \mypsr\ based on the fact that this pulse had 21 times the mean pulse energy and that \citet{RJ01} detected pulses at similar phase.  With our phase convention, this corresponds to phase 0.57, which is consistent with the phase of the second X-ray peak.

Given the very large spin-down luminosity of \mypsr, \citet{VenterThesis} proposed this MSP as a potential very-high-energy target for H.E.S.S. \citep[see also][]{FR05}. The expected spectrum was very geometry-dependent, but some flux above 100 GeV would have been expected in a screened polar cap model for an optimistic geometry. The measured $E_{\rm C}$ and the gamma-ray light curve shape presented in Figure \ref{fig:lc} disfavor this model for \mypsr.

\begin{deluxetable}{l c c}
\tablecaption{Gamma-ray and X-ray Pulse Shape Parameters of \mypsr\ \label{lcfits}}
\tablewidth{0pt}
\tablehead{\colhead{Parameter} & \colhead{Gamma-ray} & \colhead{X-ray}}
\startdata
$\Phi_{1}$ & 0.05$\pm$0.02 & 0.016$\pm$0.001\\
$\sigma_{1}$ & 0.09$\pm$0.02 & 0.013$\pm$0.001\\
$\Phi_{2}$ & 0.46$\pm$0.01 & 0.564$\pm$0.001\\
$\sigma_{2}$ & 0.05$\pm$0.01 & 0.023$\pm$0.001\\
$\Delta$ & 0.41$\pm$0.02 & 0.547$\pm$0.001\\
\enddata
\tablecomments{Peak positions are given by $\Phi_{1}$ and $\Phi_{2}$ with widths $\sigma_{1}$ and $\sigma_{2}$ (standard deviations) for the first and second peaks, respectively.  All peaks are fit with Gaussians.  The last row reports the phase separation ($\Delta$) between the first and second peaks in each waveband.}
\end{deluxetable}
\clearpage

\section{DISCUSSION}\label{disc}
\subsection{MULTI-WAVELENGTH LIGHT CURVES}\label{mwlc}
The relative phasing of the multi-wavelength light curve components in Figure \ref{fig:lc} presents a challenge to pulsar emission models.  Our preliminary attempts to explain the gamma-ray and radio light curves of \mypsr\ using geometric models yielded the following general conclusions.

It is extremely difficult, if at all possible, to obtain three radio peaks of the correct shape and position in phase by invoking only a single radio cone per magnetic pole \citep[e.g.,][]{Story07}. If instead one attempts to model the first and third radio peaks as originating from opposite magnetic poles, an interpretation supported by the 0.35 GHz profile, the chosen value of the observer angle ($\zeta$) has to be within $\sim4^{\circ}$ of 90\DG\ with a magnetic inclination angle ($\chi$)  between 40\DG\ (required so that both P1 and P3 would be visible) and 60\DG (to provide the correct radio peak multiplicity). This geometry results in the correct radio phase separation but cannot produce the correct gamma-ray peak positions (and shapes in some cases) when using standard, geometric realizations of outer-magnetospheric emission models \citep[e.g.,][]{OG,TPC}.  Stated in a different way, one may find reasonable gamma-ray profile fits (e.g., at $\chi=40^{\circ}$ and $\zeta=85^{\circ}$, although the peak separation is somewhat small and we have to choose a different fiducial phase), but then the radio peak multiplicity and / or peak positions are not correct.  There is therefore a tension between the gamma-ray and radio profiles in terms of the most preferred fit.

It is also possible to model the first two radio peaks using a radio cone above a single pole.  This interpretation would be consistent with polarization measurements indicating high linear and low circular polarization as well as a nearly constant position angle in these peaks \citep[indicative of non-caustic, conal emission,][]{BackerSallmen97,Stairs99}.  The third peak may arise from the opposite pole.  However, this is problematic when using the standard prescription for radio emission height \citep[e.g.,][]{KG03,Story07}.  The maximum peak separation for the radio P1 and P2 is obtained when $\chi\sim\zeta$ (i.e., a small impact angle), and matching the observed peak separation requires $\chi$ and $\zeta$ to be $\lesssim25^{\circ}$, which does not reproduce the observed gamma-ray profile well and predicts roughly symmetric radio peaks, contrary to the data.  On the other hand, choosing a large $\chi$ and $\zeta$ to more closely match the gamma-ray profile leads to too small a radio peak separation.  \citet{BackerSallmen97} attempted to fit the polarization position-angle swing of \mypsr\ under this assumption but were unable to match the gradient across P1.  Assuming that P1 and P3 were from opposite poles and P2 was a distant conal component from the same pole as P1, \citet{BackerSallmen97} found a reasonable fit to the polarization position-angle swing of \mypsr\ with $\chi=50^{\circ}$ and $\zeta=90^{\circ}$.  Such a solution gives the correct phasing for P1 and P3, but cannot reproduce the radio or gamma-ray profile shapes in the context of the above emission models.

Alternatively, \citet{Venter12} predicted that this pulsar may plausibly have (some) aligned gamma-ray, X-ray, and radio peaks based on the near alignment of the first X-ray and radio peaks.  In fact, a subset of gamma-ray MSPs exists in which the radio and gamma-ray peaks occur at nearly the same phase \citep{LATJ0034,Freire11,Guillemot12,Espinoza13}; however, while the first radio and gamma-ray peaks are nearly aligned and the second gamma-ray peak is nearly aligned with the third radio peak, no gamma-ray feature matches the second radio peak, which is not visible at 0.35 GHz.  In this sense, \mypsr\ is similar to PSR B1957+20 for which the two peaks in the 0.3 GHz pulse profile both have counterparts in the gamma-ray light curve but the additional component at 1.4 GHz, which occurs between the two lower-frequency peaks, does not \citep[first noted by][]{Espinoza13}.  When comparing to the 0.8 GHz radio profile presented by \citet{Rots98} we note that this peak is less prominent at lower frequency.  The radio spectral indices of MSPs with aligned radio and gamma-ray peaks tend to be softer than other gamma-ray MSPs \citep{Espinoza13}; with a spectral index of $\sim-2.4$ \citep{Lyne87} \mypsr\ could plausibly belong to this subset of gamma-ray MSPs.

A possible explanation for the near alignment of the first gamma-ray, X-ray and radio peaks and the second gamma-ray peak with the radio P3 is that they are all caustic peaks formed in the outer magnetosphere due to relativistic effects. \citet{BackerSallmen97} discussed such a model for the radio emission assuming that P2 was a polar cap beam while P1 and P3 came from the outer-gap region.  In such a model assuming co-located emission regions \citep{Venter12} the small phase differences of the first peaks in all wavebands may be reproduced by invoking slightly offset emission altitude ranges (constrained by the peak shapes). The phase difference between the second gamma-ray peak and third radio peak may be similarly explained.  In this case then, the radio P2 could come from nearer the polar cap, since it occurs at the phase expected for one of the magnetic poles.  It is not clear if shifted altitude ranges could explain the larger offset between the second gamma-ray and X-ray peaks.  Also, it would be difficult to model both the gamma-ray peaks and the radio P1 and P2 using altitude-limited models, given the relative phase lags between these peaks.  For a low-altitude geometry, the position of the second radio peak may indeed be reproduced, but then it is very difficult to reproduce the actual position of the first radio peak given the fact that the radio emitting region cannot be too extended, or it would yield peaks that are much too broad.  A caustic origin, in the outer magnetosphere, for the non-thermal X-ray emission could also plausibly explain both the pulsed and unpulsed component as noted by \citet{Bogdanov11}.  Modeling the actual pulse shapes across all wavebands will be difficult and this scenario may be in conflict with expectations from the polarization data \citep[aligned MSPs typically have no observed radio polarization][]{Venter12,Espinoza13}.  Clearly, understanding the nature of the multi-wavelength light curves of \mypsr\ will require moving beyond the standard assumptions (e.g., fine tuning the azimuthal dependence of the emissivity of high-altitude caustic radio emission) about radio and gamma-ray emission geometries.

\subsection{LUMINOSITY}\label{lum}
The gamma-ray luminosity of \mypsr\ can be calculated as $L_{\gamma} = 4\pi f_{\Omega} G d^{2}$, where $f_{\Omega}$ is a geometric correction factor accounting for the fact that the pulsar emission is not isotropic and is typically $\sim$1 for outer-magnetospheric emission models \citep{Watters09,Venter09}.  Using this formula and the results of the phase-averaged analysis with the additional off-peak source, we calculate $L_{\gamma}/f_{\Omega} =(4.0\pm1.0)\times 10^{34}$ erg s$^{-1}$.  Assuming $f_{\Omega} = 1$, we calculate the efficiency with which rotational energy is turned into HE gamma rays to be $\eta_{\gamma}\equiv L_{\gamma}/\dot{E} =0.018\pm0.005$.

\citet{Foster88} noted that the period of \mypsr\ is nearly a factor of 2 smaller than the theoretical minimum assuming a mass of 1.4 $M_{\odot}$ and accretion at the Eddington limit.  The minimum period they derive depends on the pulsar's surface magnetic field (which is derived from \Pd), mass, and radius \citep[e.g.,][]{Alpar82,Verbunt87} as well as on models of accretion by neutron stars \citep[e.g.,][]{vdH77,GL79} which could be uncertain by 50\%.   This discrepancy may imply either a more massive neutron star, super-Eddington accretion, or that the observed \Pd\ is artificially increased by the gravitational acceleration field in the cluster along our line of sight (as given by Eq.~\ref{pdotInc} where $a_{l}$ is the line-of-sight acceleration):

\begin{equation}\label{pdotInc}
\Big(\frac{\dot{P}_{\rm obs}}{P}\Big)\ =\ \Big(\frac{\dot{P}}{P}\Big) + \frac{a_{l}}{c}.
\end{equation}

The latter explanation was deemed unlikely by \citet{Foster88}, and \citet{Phinney93} showed that the maximum $|a_{l}|$ for M28 was $9\times10^{-9}$ m s$^{-2}$, which suggests that $\leq$6.6\% of the observed \Pd\ is not intrinsic.  Using Eq.~6 in the appendix of \citet{Freire05} and the central velocity dispersion parameters from the Harris catalog\footnote{\url{physwww.physics.mcmaster.ca/~harris/mwgc.dat}} \citep{Harris96} and the distance of M28, we find a slightly higher maximum $|a_{l}|$ of $2\times10^{-8}$ m s$^{-2}$.  However, this still suggests that, at most, only 14\% of the observed \Pd\ of \mypsr\ is not intrinsic.  We can use $\eta_{\gamma}$ to assess the need for any line-of-sight acceleration contribution to $\dot{ P}_{\rm obs}$.  The average $\eta_{\gamma}$ for pulsars with \Ed$\in[0.4,4]\times10^{36}$ erg s$^{-1}$ in the second LAT catalog of gamma-ray pulsars, excluding those pulsars with no distance estimate or with distance uncertainties leading to systematic uncertainties on $\eta_{\gamma}$ of more than 50\%, is 0.116 with a large spread (RMS = 0.090) \citep{2PC}.  While the value of $\eta_{\gamma}$ we calculate is somewhat below the average, it is not uncommon in this \Ed\ range; in particular, out of the 16 pulsars we use for this average 4 (25\%) have $\eta_{\gamma}<0.02$.  Thus, we see no strong indication from $\eta_{\gamma}$ that the measured \Pd\ is significantly enhanced by the cluster potential, supporting the findings of \citet{Phinney93} that the observed \Pd\ of \mypsr\ is nearly 100\% intrinsic.  This differs from the conclusion of \citet{wuM28A} but we note that they compared results for \mypsr\ to those of MSPs in the Galactic field that have significantly lower values of \Ed\ and thus are not expected to have similar efficiencies.

\subsection{MSP POPULATION IN M28}\label{pop}
Assuming that the off-peak emission discussed in Section \ref{spec} is in fact from other pulsars in M28, despite the positional offset, and following the prescription of \citet{GCpop} we can estimate the number of energetic MSPs in M28 as,
\begin{equation}\label{Nmsp}
N_{\rm MSP}\ =\ \frac{L_{\gamma,\rm{off}}}{\langle\dot{E}\rangle\langle\eta_{\gamma,\rm{MSP}}\rangle}.
\end{equation}
\noindent{}Using the off-peak luminosity, $L_{\gamma,\rm{off}}=(9.4\pm2.0)\times10^{34}$ erg s$^{-1}$; average \Ed\ of MSPs in globular clusters, $\langle\dot{E}\rangle = (1.8 \pm 0.7) \times 10^{34}$ erg s$^{-1}$ \citep{47Tuc}; and average MSP gamma-ray efficiency, $\langle\eta_{\gamma,\rm{MSP}}\rangle = 0.245$ calculated from \citet{2PC} excluding 10 MSPs for which the distance uncertainties lead to systematic uncertainties on $\eta_{\gamma}$ greater than 50\% and one with an unrealistic $\eta_{\gamma}>1$; we calculate $N_{\rm MSP} = 20\pm9$ for M28, not counting \mypsr.  We note that this value is highly dependent on the value of $\langle\eta_{\gamma,\rm{MSP}}\rangle$ chosen and thus the systematic uncertainty of this estimate is greater than the statistical value we quote.

If the off-peak emission is in fact not associated with M28, we can use the upper limit calculated at the cluster position in the off-peak interval to limit $N_{\rm MSP} \leq 5$, not including \mypsr.  This is less than the number of pulsars known in M28, but is also highly dependent on the value of $\langle\eta_{\gamma,\rm{MSP}}\rangle$ used, as noted previously.  Therefore, from this upper limit we can say only that there is no strong evidence for many pulsars in M28 beyond those already known.

We can make another estimate of the gamma-ray flux contributed by the other pulsars in M28 if we statistically correct the ${\dot P}$ of the other known pulsars in M28 for the effect of $a_{l}$.  While we do not know $a_{l}$ for the individual pulsars, we can estimate the maximum acceleration at the projected distance from the cluster core and compute the probability distribution of $a_{l}$ following \citet{Phinney93}.  Using the known projected distances of each object, this gives us a probability distribution for intrinsic spin-down rate of each pulsar, solving Eq.~\ref{pdotInc} for $\dot{P}$, and hence the intrinsic spin-down luminosity.

Using a King-type cluster model with pulsar density $n_{\rm PSR} \propto r^{-3/2}$, a simple $L_{\gamma} \propto \sqrt{\dot{E}}$ efficiency law, and assuming the off-peak emission is associated with M28 we estimate that \mypsr\ should contribute 0.33$\pm$0.05 of the combined gamma-ray energy flux of the 12 known pulsars in M28.  This agrees well with the observed ratio of the phase-averaged energy fluxes with and without the additional off-peak source of $0.34\pm0.06$.  This analysis suggests that the other known pulsars in M28 easily provide enough luminosity to account for the off-peak emission.  In turn, this implies that the number of energetic pulsars in M28 may not be much larger than 12 and that MSP radio beams cover a large fraction of the sky, comparable to that of the gamma-ray beams.  It also suggests that the next brightest pulsar (likely C, I, or K) could provide as much as $\sim 1/4$ the gamma-ray flux of \mypsr.

The high incidence (5/12, after correcting for $a_{l}$) of $\dot{E}>10^{35}$ erg s$^{-1}$ MSPs in M28 implies that not so many unknown pulsars need to contribute to the unpulsed flux, unless they are much fainter in gamma rays than \mypsr.  Though lower, this estimate does agree with the value of $N_{\rm MSP}=20\pm9$ MSPs using Eq.~\ref{Nmsp}.  Our first estimate relies on comparison with the average $\eta_{\gamma}$ of nearby field MSPs with typical $\dot{E} \sim 10^{34}$ erg s$^{-1}$ while this last estimate relies on the simple $L_{\gamma} \propto \sqrt{\dot{E}}$ scaling.  It is likely that the true pulsar efficiency at very low \Ed\ departs from this law \citep[e.g.,][]{Harding02,Zhang04,TWC10}.

While our analysis indicates that magnetospheric emission from the other known pulsars in M28 can plausibly account for the off-peak emission, eight of these pulsars are in binary systems, two are observed to eclipse and three are estimated to have low-mass ($\lesssim 0.02 M_{\odot}$) companions.  Shocked emission from interactions between the pulsar wind and the companion stars in these systems may contribute to the emission observed by the LAT \citep{HG90,Takata12}.  The classic example of such emission is PSR B1259$-$63 \citep{AbdoB1259} from which unpulsed GeV emission is only detected near periastron.  However, searches for orbitally-modulated emission from energetic gamma-ray MSPs have resulted in no firm detections \citep{Guillemot12,Pletsch12} with the best evidence, to date, a 2.3$\sigma$ indication of orbital modulation above 2.7 GeV from PSR B1957+20 \citep{WuB1957} and a 2$\sigma$ indication for PSR J0610$-$2100 above 3 GeV \citep{Espinoza13}.  Thus, any non-magnetospheric emission from the known energetic binary MSPs in M28 is not expected to be strong and should not affect our previous conclusions.  However, we did fold the data at the orbital periods of the M28 pulsars in binary systems and found no significant signal.

\section{CONCLUSIONS}\label{conc}
\mypsr\ is the second MSP located in a globular cluster from which significant gamma-ray pulsations have been detected.  Similar to PSR J1823$-$3021A, the derived efficiency of \mypsr\ supports previous assertions that the observed \Pd\ is largely intrinsic, providing further evidence that this is an unusually energetic MSP.  This is further highlighted by other properties of \mypsr\ (such as the giant radio pulses and HE emission) that are generally observed in young, very energetic, and fast-spinning pulsars.

\mypsr\ and PSR J1823$-$3021A have \Pd\ values $\sim100$ times larger than typical of other MSPs with comparable spin periods, which implies that their lives as MSPs will be $\sim100$ times shorter -- a few tens of millions of years.  This means that these pulsars must be forming at a rate comparable to that of other MSPs in globular clusters, which are $\sim100$ times more numerous but also $\sim100$ times longer lived.  It is not clear whether these energetic MSPs formed by the same processes that formed the more normal MSPs; or by some alternative process \citep[e.g.,][]{Ivanova08}.  If the formation process is the same, then they do not represent a separate population and are part of the same continuum.  This would indicate that the `normal' formation mechanism is able to produce MSPs with a wider range of magnetic fields than is typically assumed.  This would also imply that such very energetic MSPs should be observed in the Galaxy outside of globular clusters.  To date, the only such field MSP that might belong to this class is PSR B1937+21.  If no pulsars like \mypsr\ and PSR J1823$-$3021A are found in the Galaxy, that would lend credence to the hypothesis that these two MSPs are part of a separate population that forms only in globular clusters or other environments with very high stellar density.  \citet{VF13} note that all ``young'' pulsars in globular clusters are found only in clusters with a high rate of stellar encounters \emph{per binary}, where there is a reasonable chance of X-ray binaries being disrupted during recyling.  This may be one way to explain why both \mypsr\ and PSR J1823$-$3021 are isolated without invoking alternate formation scenarios.  Only improved statistics, from new MSP discoveries in globular clusters and the Galactic field, will tell.

The multi-wavelength light curves of \mypsr\ suggest a complex relationship between the different emission regions.  The fist gamma-ray and X-ray peaks (and possibly the second X-ray peak) are consistent with the phase of giant radio pulses.  While the association of the off-peak emission with M28 is unclear, we find no strong evidence, in any case, that the population of energetic pulsars is much larger than the 12 pulsars already known.

Multi-wavelength models of globular cluster spectra have different assumptions on the origin of the HE emission and create different expectations for the spectral shape. In the case where the HE emission results from the cumulative pulsed curvature radiation from MSPs, an additional unpulsed inverse-Compton component may dominate in the TeV band \citep[e.g.,][]{ZBR13,Kopp13}. This second component is expected to be much lower and would largely leave the curvature radiation signature unaffected, consistent with the observed spectrum that cuts off at several GeV and detection of gamma-ray pulsations from two globular cluster MSPs. Conversely, if the HE emission is due to inverse-Compton scattering \citep{Cheng10} the spectral shape may mimic a curvature radiation spectrum in the GeV range, sometimes also predicting TeV spectral components, for some parameter choices.

This detection was enhanced by the use of LAT data that have been reprocessed with improved instrument calibration constants and demonstrates that, as the \fermi\ mission continues, improvements in the data reconstruction and analysis methods will continue to enhance LAT science.

\acknowledgments

\begin{center}
\emph{ACKNOWLEDGMENTS}
\end{center}

The \fermi\ LAT Collaboration acknowledges generous ongoing support from a number of agencies and institutes that have supported both the development and the operation of the LAT as well as scientific data analysis.  These include the National Aeronautics and Space Administration and the Department of Energy in the United States, the Commissariat \`a l'Energie Atomique and the Centre National de la Recherche Scientifique / Institut National de Physique Nucl\'eaire et de Physique des Particules in France, the Agenzia Spaziale Italiana and the Istituto Nazionale di Fisica Nucleare in Italy, the Ministry of Education, Culture, Sports, Science and Technology (MEXT), High Energy Accelerator Research Organization (KEK) and Japan Aerospace Exploration Agency (JAXA) in Japan, and the K.~A.~Wallenberg Foundation, the Swedish Research Council and the Swedish National Space Board in Sweden.

Additional support for science analysis during the operations phase is gratefully acknowledged from the Istituto Nazionale di Astrofisica in Italy and the Centre National d'\'Etudes Spatiales in France.

The Nan\c{c}ay Radio Observatory is operated by the Paris Observatory, associated with the French Centre National de la Recherche Scientifique (CNRS).

The Westerbork Synthesis Radio Telescope is operated by Netherlands Foundation for Radio Astronomy, ASTRON.

The Lovell Telescope is owned and operated by the University of Manchester as part of the Jodrell Bank Centre for Astrophysics with support from the Science and Technology Facilities Council of the United Kingdom.

Pulsar research at the University of British Columbia is supported by a NSERC Discovery Grant and by the Canada Foundation for Innovation.

Portions of this research performed at the Naval Research Laboratory are sponsored by NASA DPR S-15633-Y.

\textit{Facilities:} \facility{Fermi(LAT)}, \facility{NRT}, \facility{WSRT}, \facility{Lovell}

\bibliographystyle{apj}
\bibliography{PSRB1821}

\end{document}